\documentclass[epj]{aa}
\usepackage{psfig}
\usepackage{epsfig}
\usepackage{citesort}       

\def\text#1{{#1}}

\def\lsim{\raise0.3ex\hbox{$\;<$\kern-0.75em\raise-1.1ex\hbox{$\sim\;$}}}
\def\gsim{\raise0.3ex\hbox{$\;>$\kern-0.75em\raise-1.1ex\hbox{$\sim\;$}}}

\def\MeV{{\rm MeV}} 
\def\eV{{\rm eV}}

\def\npbps#1#2#3{  { Nucl. Phys. }(Proc. Suppl.){\bf B #1} (19#2) #3} 
\def\plb#1#2#3{    { Phys. Lett. }{\bf B #1} (19#2) #3}

\def\prl#1#2#3{    { Phys. Rev. Lett. }{\bf #1} (19#2) #3}

\begin{document}
\title{Statistically improved Analysis of Neutrino Oscillation Data with the latest KamLAND results}
\author{ V. Antonelli$^{a,b}$, 
E. Torrente-Lujan$^{c}$\\
{\small\sl
$^{a}$ Dip. di Fisica, Universit\`a degli Studi di Milano, Milano, Italy\\ 
$^{b}$ I.N.F.N., Sezione di Milano, Milano, Italy\\
$^{c}$ Dept. de Fisica, Universidad de Murcia, Murcia, Spain\\
\email{vito.antonelli@mi.infn.it, torrente@cern.ch}}
}

\titlerunning{Analysis of Neutrino Oscillations...}
\authorrunning{Antonelli, Torrente-Lujan}

\date{}

 \abstract{
We present an updated analysis of all available solar and 
reactor neutrino data, emphasizing in particular the totality of the KamLAND (314d live time) 
results and including for the first time the solar $SNO$ 
(391d live time, phase II NaCl-enhanced)  spectrum data.
As a novelty of the statistical analysis, we study 
the variability
 of the KamLand results with respect the use 
of diverse statistics. A  new statistic, 
not used before is proposed.
Moreover, in the analysis of the SNO spectrum a novel technique 
is used in order to include full correlated errors among bins.
 Combining all data, we obtain the following best-fit parameters: 
we determine individual neutrino mixing parameters and 
their errors
$ \Delta m^2= 8.2\pm 0.08\times 10^{-5} \eV^2,\quad 
\tan^2\theta=  0.50^{+0.12}_{-0.07}.$ 
The impact of these results is
discussed.
We also estimate the individual elements of the  neutrino 
mass matrix.
In the framework of three neutrino oscillations we obtain the 
mass matrix:
{\small
\begin{eqnarray}
M&=& eV 
\pmatrix{ 1.0+
4.0\pm 3.2\ 10^{-5}& 4.2\pm 3.2\ 10^{-5} &-13.5\pm 14.0\ 10^{-5} 
\cr  
4.2\pm 3.2\ 10^{-5}& 1.0+4.3\pm 3.5\ 10^{-5} &-13.5\pm 14.5\ 10^{-5} 
\cr 
13.5\pm 14.0\ 10^{-5}& -13.5\pm 14.5\ 10^{-5} &1.0+100.0\pm 30.0\ 10^{-5} 
}.
\nonumber
\end{eqnarray}
}
}



\maketitle

\section{Introduction}

Evidence of antineutrino disappearance in a beam of antineutrinos  
in the Kamland experiment has been presented \cite{klJun}. 
The analysis of previous experimental results on reactor physics and solar 
neutrinos \cite{klothers} in terms of neutrino oscillations 
has largely improved our knowledge of neutrino mixing. 
Thus, the solar neutrino data evidence prior to autumn of 2003 
 converged to two 
distinct allowed regions in  parameter space, often referred to as LMAI 
(centered around the best-fit point of
$\Delta m^2_{\odot}=7.1\times10^{-5} \eV^2$,
$\tan^2\theta_{\odot}=0.47$)
and LMAII (centered around
$\Delta m^2_{\odot}=1.5\times 10^{-4} \eV^2$, 
$\tan^2\theta_{\odot}=0.48$).
The inclusion of SNO phase II  data 
 eliminated the LMAII region at about $4\sigma$.

 The KamLAND  
measurementes presented recently \cite{klJun} 
corresponds to the first about 317 days live time, they 
confirm the conclusions obtained from previous data and give a 
more stringent limit on the neutrino mass parameters.  
As we will see in this work, the new KamLAND
information further excludes the LMAII region now at 
$\approx$ 5$\sigma$.

In addition to an increased statistics,
a significant change in the KamLAND analysis technique is
 related to the fiducial volume definition.
Whereas in the previous setup,
 events taking place at the outer edge of
the nylon balloon were rejected \cite{klJun,Eguchi:2002dm}, the recent 
analysis adopts a more
sophisticated coincidence-measurement technique to exclude unwanted
backgrounds. 
Additionally,
a better understanding of the fuel cycle on the reactors has lead to 
the collaboration 
to estimate
the incoming neutrino
flux with better accuracy: 
the estimated error on this initial flux 
$\phi_0\left(\overline{\nu}_{e}\right)$
is now of the order of  2\%.

The aim of this work is to present a comprehensive updated 
analysis of 
all recent solar neutrino data including the KamLAND 
reactor-experiment results to determine the extent of the remaining 
viable region in the parameter space and to 
obtain favoured values for the neutrino physical parameters 
in a two-neutrino framework, and, with the inclusion of results 
coming from atmospheric oscillation evidence give an estimation 
of the elements of the three neutrino mass matrix. 
Some key analysis novelties are presented along this work, 
for 
example regarding the treatment of sistematic correlations in 
the analysis of the SNO spectrum (day+night) and a improved analysis of the 
KL data with an appropriate  consideration of its low statistics 
data bins.

The structure of this paper is the following: in
section~{\ref{kamland}} discuss our approach to the latest KamLAND
results and older solar evidence. 
All solar neutrino experiments are discussed in
section~{\ref{solar}}. We specially 
discuss the importance of the SNO data
and the spectrum results.
We then proceed, in section~{\ref{analysis}}, 
to explain the procedure adopted in our 
analysis with some emphasys on the special 
treatment needed by the KamLand and SNO spectra 
and in section~{\ref{results}} we present our results. 
Finally 
we summarize and conclude in section~{\ref{summary}}.

\section{The KamLAND and solar evidence} 

\label{kamland} 
\label{simulations}

Reactor anti-neutrinos with energies above 1.8 MeV produced in some 53 
commercial reactors are detected in the KamLAND detector via the inverse 
$\beta$-decay reaction $\overline{\nu}_{e}+p\rightarrow n + e^{+}$. The mean 
reactor-detector distance and energy window of these $\overline{\nu}_{e}$ 
makes KamLAND an ideal testing ground for the LMA region of the $\nu_{\odot}$ 
parameter space. The first results published by the KamLAND collaboration 
eliminated all possible solutions to the solar neutrino problem (SNP) except 
the LMA region of the parameter space \cite{Eguchi:2002dm}. The 
sensitivity of this experiment to the $\Delta m^2$ parameter divided 
the previously whole LMA region into two distinct regions, the one 
relative to the smaller mass-squared difference being preferred by 
data \cite{postKamland}.

In our analysis (see also Ref.\cite{torrente} for further 
details) 
we  model the reactor incoming flux 
 through a  constant, time-averaged fuel 
composition for all of the commercial reactors within detectable 
distance of the Kamioka site, namely  
${}^{235}\text{U}=56.3\% $, 
${}^{238}\text{U}=7.9\% $, 
${}^{239}\text{Pu}=30.1\% $, and 
${}^{241}\text{Pu}=5.7\% $. 
We used the full cross-section including 
electron recoil corrections. We analyzed the data above   
threshold of $2.6 \MeV$, as the low-energy end of the 
spectrum had effectively no events. 
The information relative
to the no-oscillation initial flux for the two periods can be
extracted from fig. (1.a) of \cite{klJun}.
We neglect all backgrounds, 
including geological background above 2.6 MeV.  
We use the resolutions published by the collaboration 
for the two different 
data sets, namely 
$\sigma\left(E\right) = 6.2\%/\sqrt{E}$ for 
the recent data (post upgrade)  
and $\sigma\left(E\right) = 7.3\%/\sqrt{E}$ for earlier 
data (pre-upgrade).
The total systematic error is estimated at $6.5 \%$. 

In order to use all the data available, we use a simple 
MC simulation  to estimate an 
equivalent efficiency for the two pre-upgrade and post-upgrade
phases. 
Finally, no matter effects are taken into consideration for the KamLAND data
alone, as it was shown that, for this experiment,
any asymmetry due to matter effects is negligible for $\Delta m^2$ of
the order of $10 \times 10^{-5}$ eV$^2$.

\subsection{Solar data} 
\label{solar} 

The most ponderous data present in our analysis come from the solar 
neutrino experiments. The experimental results are compared to an 
expected signal which we compute numerically by 
convoluting 
solar neutrino 
fluxes \cite{bpb2001}, sun  and earth 
oscillation probabilities, neutrino cross sections 
and detector energy response functions. We closely follow 
the same methods already well explained in previous works
\cite{Aliani:2001zi,Aliani:2001ba,Aliani:2002ma,Aliani:2002rv}, 
we will mention here only a few aspects 
of this computation.
We determine the neutrino oscillation probabilities 
using the standard
methods found in literature~\cite{torrente}, as explained 
in detail in~\cite{Aliani:2001zi} and in~\cite{Aliani:2002ma}. 
We use a thoroughly numerical method to calculate the 
neutrino evolution equations in the presence of matter for all 
the parameter space.
For the solar neutrino case 
the calculation is split in three steps, corresponding to 
the neutrino propagation inside the Sun, in the vacuum 
(where the propagation is computed analytically) and in the Earth.
We average over the neutrino production point inside the Sun 
and we take the electron number density $n_e$ in the Sun by the BPB2001 model~\cite{bpb2001}.
The averaging over the annual  variation of the orbit
is also exactly performed.
To take the Earth matter effects into account, we adopt a
 spherical model of the Earth  density and chemical 
composition~\cite{torrente}.
The joining of the neutrino propagation in the three different 
regions is performed exactly using an evolution operator
formalism~\cite{torrente}.
The final survival probabilities are obtained from the 
corresponding (non-pure) density matrices built from 
the evolution operators in  each of these three regions.

In summary,
double-binned day-night and zenith angle bins are 
computed in order to analyze the full SuperKamiokande data 
\cite{Smy:2002fs}, whereas single-binned data is used for the SNO 
detector \cite{newSNO,Poon:2001ee,Ahmad:2002jz}. The global signals
only are used for the
radiochemical experiments Homestake \cite{homestake}, SAGE 
\cite{sage1999,sage}, GallEx \cite{gallex} and GNO \cite{gno2000}. 
The next paragraphs are dedicated to a description of 
 SNO component of the solar evidence.

\label{NaCl} 

The Sudbury Neutrino Observatory (SNO) collaboration has 
presented data relative to the NaCl phase of the experiment 
\cite{newSNO,newSNO03}. 
We remind that the addition of  
NaCl to a pure $\text{D}_{2}\text{O}$ detection medium has the effect 
of increasing the detector's sensitivity to the neutral-current (NC) 
reactions within its fiducial volume. The NC detection efficiency has 
changed from a previous 'no-salt' phase of approximately a factor 
three.
This and other novelties have made it possible for the SNO 
collaboration to analyze their data without making use of the 
no-spectrum-distortion hypothesis. Furthermore, they have 
adopted a new 'event
topology' criterion \cite{newSNO03} 
to distinguish among the different channels within 
the detector.  
The SNO Collaboration has now devised a new data-analysis 
technique which relies on the topology of the three different 
events. The new 
parameter ($\beta_{\ell}$) relative to which they marginalize is known 
as the 'isotropy' of the Cerenkov light distribution was used to 
separate the CC, ES and NC signals, something that was not possible in 
th previous two data sets. The measured fluxes are reported in 
\cite{newSNO}.

The comparison of the new SNO results and the previous 
phase-II data can easily be made because the SNO collaboration 
has included in the 
recent paper results which were obtained following their previous 
method, along with the new unconstrained data. The new 
results are 
compatible with the previous ones. It seems that the overall 
effect of un-constraining the analysis is an increase in 
the measured fluxes, although the estimated total 
$\Phi_{\text{B}}$ has decreased relative to the previous 
best-fit value, leaving even less space for eventual sterile 
neutrinos. We make use of all day+night 
published data and refer the 
reader to section~{\ref{analysis}} for details.
The procedure we used to introduce the SNO spectrum data is an
extention of the one use for the SK spectrum analysis and is 
explained in the next section.

\section{Our analysis} 
\label{analysis} 

We use  standard statistical techniques to test the non-oscillation 
hypothesis. Two different sets of analyses are possible with the 
present data on neutrino oscillations:  
1) short-baseline reactor data, solar data including the SK spectrum  
and previous phase-I (CC only) SNO spectrum, phase-II SNO global
result, combined with new the KamLAND spectrum and, 
2) the previous set with the use of the phase-II SNO spectrum result
and the new KamLAND data.
In order to use all the SNO data, we consider the phase-I and phase-II
results as two distinct but fully correlated experiments.  

For the purpose of this analysis, a  $\chi^{2}$ function 
is defined which is the 
the sum of the distinct contributions. The 
contribution of all solar
neutrino experiments  
is summarized in the term: 
 \begin{equation} 
\chi^{2}_{sun}=\chi_{rad}+\chi_{SK}^{2}+\chi^2_{\text{SNO}}.
\end{equation} 
Where, the  $\chi^{2}$ function for the global rates 
of the radiochemical 
experiments is as follows,
\begin{equation}\label{chi_radiochemical} 
\chi^{2}_{\text{rad}}=\left(\mathbf{R}^{\text{th}}-\mathbf{R}^{\text{exp}}\right)^{T}\left(\sigma_{\text{sys}}+\sigma_{\text{stat}}\right)^{-1}\left(\mathbf{R}^{\text{th}}-\mathbf{R}^{\text{exp}}\right),
\end{equation} 
where $\mathbf{R}^{\text{th,exp}}$ are length-two vectors containing the  
theoretical (or experimental) signal-to-no-oscillation expectation for 
the chlorine and gallium experiments. Correlated systematic and uncorrelated 
statistical errors are considered in $\sigma_{\text{syst}}$ and 
$\sigma_{\text{stat}}$ respectively. Note that 
the parameter-dependent $R^{\text{th}}$ is an averaged day-night quantity, as 
the radiochemical experiments are not sensitive to day-night variations. 
For the next component
We consider the double-binned SK spectrum comprising of 8 energy bins for a 
total of 6 night bins and one day bin. The $\chi^{2}$ is given by 
\begin{equation}\label{chi_spectrum_sk} 
\chi^2_{\text{SK}}= (\alpha\mathbf{R}^{\text{th}} -\mathbf{R}^{\text{exp}})^T  
\left (\sigma^{2}_{\text{unc}} + \sigma^{2}_{\text{cor}}\right )^{-1} 
 (\alpha\mathbf{R}^{\text{th}}-\mathbf{R}^{\text{exp}}). 
\end{equation} 
The covariance matrix $\sigma$ is a 4-rank tensor containing 
information relative to the statistical errors and energy and 
zenith-angle bin-correlated and uncorrelated uncertainties. 
Since the publication of the first SNO NC results, we have adopted 
their estimate of $\phi_{B}$ and incorporated the new parameter 
$\alpha$ in the $\chi^2$ representing the normalization with respect 
to this quantity. In determining our best-fit points, 
we minimize with respect to it. 
Note that the quantities $\mathbf{R}^{\text{exp}}$ and 
$\mathbf{R}^{\text{th}}$ contain the number of events 
normalized to the no-oscillation scenario.

We deal next with the SNO component.
We present two  different  analysis of some of 
the phase II SNO data sets including total day/night quantities.

The first consideres the   global signal alone, 
the second incorporates the total signal  spectrum. 
We consider here the two SNO  results as if coming from two 
independent experiments, but fully correlated. 
We use the backgrounds as listed in tables 
X of Ref.\cite{newSNO}
and 
1 of 
Ref.\cite{Ahmad:2002jz} for the phase-II data. The detector 
resolution is obtained from 
Refs. \cite{newSNO,howto}. 

In the second case we make  use of the spectral data.
The spectrum used for our analusis is presented in 
 table~[\ref{sno_spectrum}]). 
The $\chi_{\text{SNO}}^{2}$ has 
the same formal expression as before, where it is 
understood that $\mathbf{R}^{\text{th,exp}}$ are now length 32 
containing two 16-bin relative to the two SNO data 
sets. We consider the two fully correlated. 

The main difficulty in using the total spectrum data lies 
in correctly estimating the, highly correlated, 
systematic error. 
By using the information contained in tables XIX and XX of 
Ref.\cite{newSNO}, we have computed the influence of all the 
different sources of error on our response function 
considering the correlation/anti-correlation as presented in 
table 1 of \cite{howto}. 
The different backgrounds spectral correlations are 
included from table XXXIV of Ref.\cite{newSNO}. 
The procedure we used to introduce the SNO spectrum data is an
extension of the one used for the SK spectrum analysis.
For each point in the parameter space $\Delta m^2$, $\theta$
we start from a correlation matrix obtained by using the 
non-deformed spectrum assumption.
We calculate, for each bin, the sum of the signals ES+NC+CC and we
extract weights for each single contributions.
After that, we compare our theoretical results with the 
ones given by
the SNO collaboration and impose a 3-$\sigma$ cut.
By using these zero-order 
weights and the correlation errors obtained
by the SNO table, we reconstruct a correlation matrix.
The correlation matrix is introduced into the $\chi^2$ analysis 
by adding a free parameter $\delta_{cor}$ which is 
determined in a minimization process togethed with the weights of
the single $i=ES+NC+CC$ contributions to the signal:
{\small
\begin{eqnarray*}
\chi^2_{SNO}&=&
\sum_{i}(\alpha R^{th}- R^{exp})^t 
(\sigma^2_{unc}+\delta_{cor}\sigma^2_{cor})^{-1}
(\alpha R^{th}-R^{exp})  \\
&&+ \chi^2_{\alpha} + \chi^2_{\delta_{cor}},
\end{eqnarray*}

}
The full process is designed to be interated a number of times,
 in practise we obtain that after two iterations the process 
 is convergent and give us the desired results.

\subsection{The Kamland statistical analysis}

The total KamLAND contribution to the $\chi^2$ is defined as:  
\begin{equation}
\label{chi_kamland} 
\chi^2_{\text{KL}} = \chi^2_{\text{KL,glob}}
+\chi^2_{\text{KL},\lambda} 
\end{equation} 
where the global contribution is simply 
 \begin{equation}
\label{chi_kamland_both} 
 \chi^2_{\text{KL,glob}}= 
 \frac{\left(R^{\text{th}}-R^{\text{exp}}\right)^2}{\sigma_{\text{stat+sys}}^2}.
\end{equation} 

The statistical consideration of the KL spectrum 
signal, $\chi^2_{KL,\lambda}$, 
is however worthy of special attention.
Due to the fact that at high energy KamLAND observes a small 
number of events alternatives should be  used instead 
the   Gaussian approximation. 
This means among other things
 that the  correlated  systematic deviations 
cannot be introduced in a straightforward way.
Due to these reasons, we use an alternative technique 
for the KamLAND data bins. 
A detailed account of some statistical considerations 
is presented in the Appendix.

It is possible to present an unified 
approach \cite{read88}
to all the commonly used  multinomial models (Pearson's, 
 log-likelihood among them) by defining  
 a family of statistics $\chi^2(\lambda)$ 
for testing the fit of 
observed frequencies $R_i^{exp}$ to the expected ones $R_i^{th}$  \cite{read88}.
All the statistics belonging to this family have 
similar well-behaved properties but however results as best 
fit parameters and exclusion regions may significantly depend on the 
use of one or another.
Any decision as to which member of the family we should use 
to finally test the null hypothesis must depend on the type 
of the departure we wish to detect.
The sensitivity of the statistic depends on how the defining 
function treats the large or small deviations.

Based on a comparative study it is recomended 
\cite{read88,stat}
to use $\chi^2(2/3)$ as a compromise candidate among the 
different  test statistics optimizing  diverse criteria as
 rate of convergence, sensitivity to the sample size and 
sensitivity to large or small 
bin deviations. The statistic corresponding to this value, the Read statistic,  is the one used in this work:
{\small
\begin{eqnarray*}
\chi^2_{KL}\left (\lambda=\frac{2}{3}\right )&=&
\frac{9}{5}
\sum_i R_i^{exp} \left ( \left( \frac{R_i^{exp}}{R_i^{th}}\right )^{2/3} - 1  
\right )\\
&&
+\frac{2}{3} \left (R_i^{th}-R_i^{exp}\right ).
\end{eqnarray*}
}

In the evaluation of $\chi^2_{\text{KL},\lambda}$ we use vectors 
that comprise therefore of 13 spectral points of width 0.425 MeV.

\section{Results and Discussion.}
\label{results}

To test a particular oscillation hypothesis against 
the parameters 
of the best fit (null hypothesis) and obtain allowed regions in 
parameter space we perform a 
 minimisation of the full function
 $\chi^2$ with respect the oscillation and the rest of ancillary 
parameters. 
A given point in the oscillation parameter space is allowed if 
 the globally subtracted quantity fulfills the condition 
 $\Delta \chi^2=\chi^2 (\Delta m^2, \theta)-\chi_{\rm min}^2<\chi^2_n(CL)$.
Where $\chi^2_(90\%,95\%,...)$ are the respective 
quantiles.
In this way we obtain best fit mass differences and angles and 
joint exclusion regions.
Additionally, we perform a second kind of analysis in order to obtain
concrete values for the individual oscillation parameters and
estimates for their uncertainties. 
We study the marginalised parameter constraints where the  $\chi^2$
quantity is converted into likelihood using the expression
${\cal L}/{\cal L}_0=$ $e^{-(\chi^2-\chi_{min}^2)/2}$.

In table~[\ref{bestfitpoints}] we report the values of the mixing parameters 
$\Delta m^2_{\odot}$, $\tan^2\theta_{\odot}$, and the $\chi^2$ 
obtained from minimization and from the peak of marginal likelihood 
distribution. 

\begin{figure*}
\centering 
\begin{tabular}{cc} 
\includegraphics[scale = 0.5]{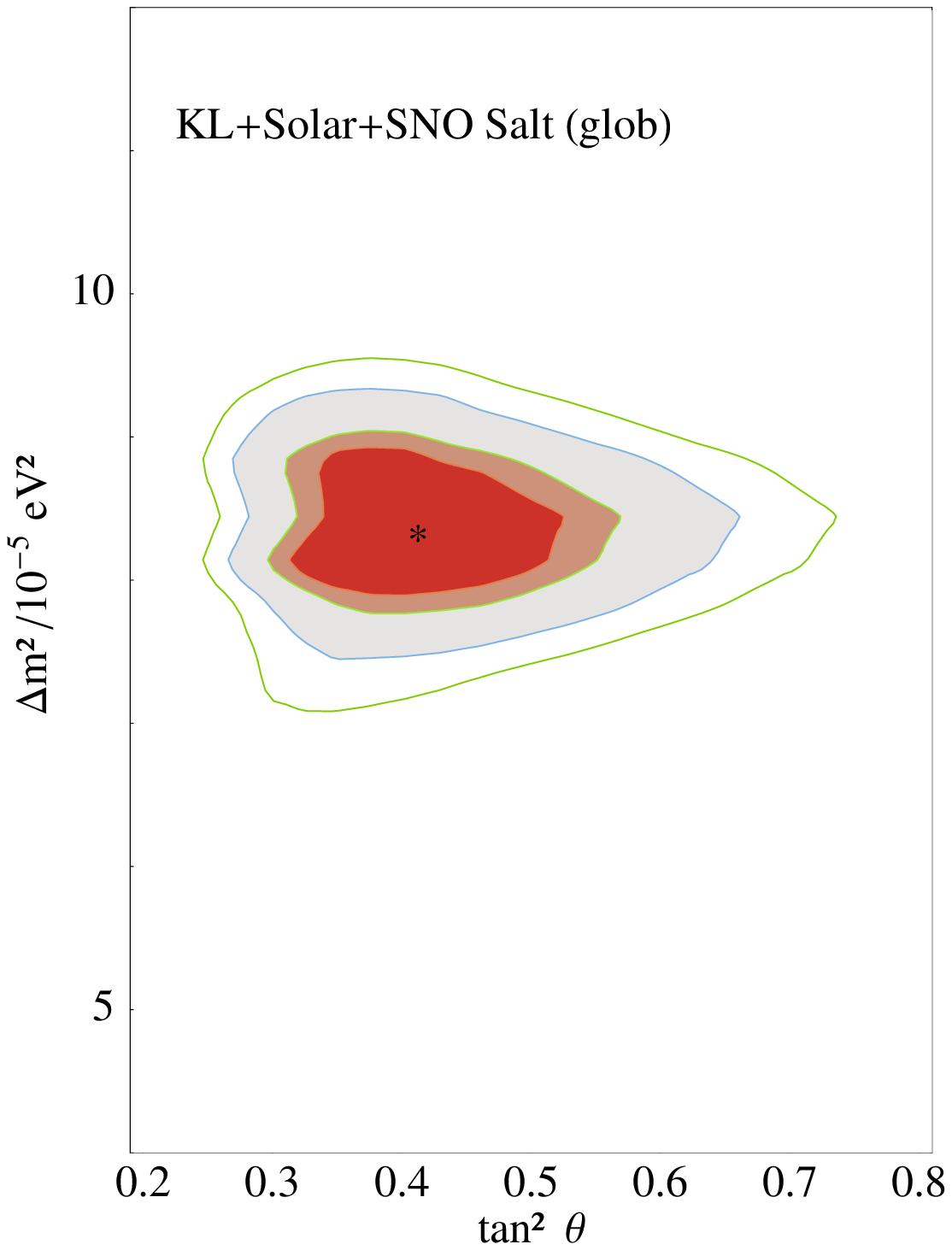}&
\hspace{-0.5cm} 
\includegraphics[scale = 0.5]{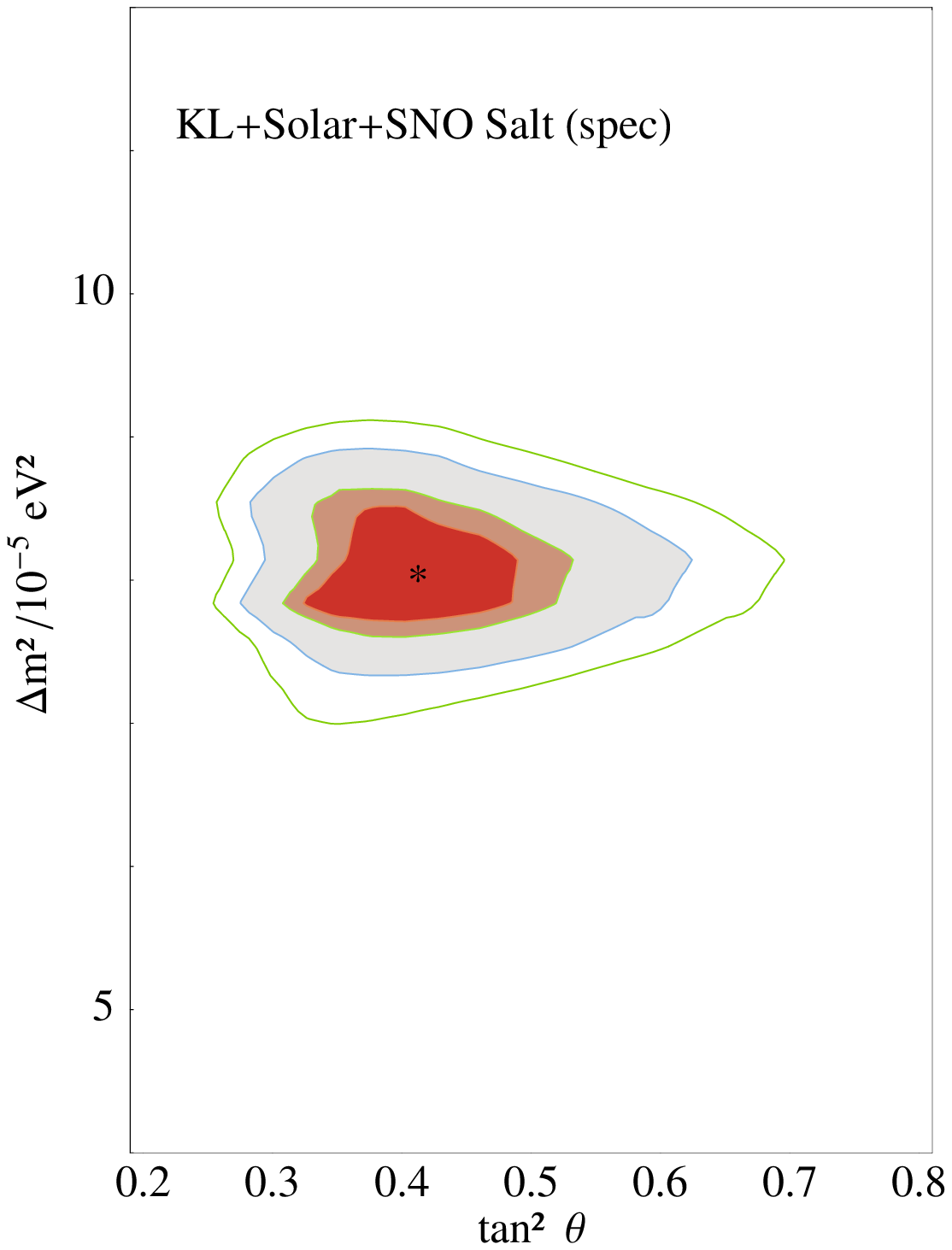} 
\end{tabular} 
\caption{   
({\it left}) Allowed region in the $(\tan^2\theta, \Delta
m_{\odot}^2)$ plane for the global analysis, which includes the
previous solar data (see e.g. \cite{ourstuff} for details) and all KamLAND Global results.  
({\it right}) Best fit solution for the spectrum analysis, including
all previous solar data,  short baseline reactor data and the KamLAND
spectrum. Best fit point given in table [\ref{bestfitpoints}].} 
\label{fexclusions} 
\end{figure*}

The results are shown in Figs.\ref{fexclusions} where we 
have generated acceptance 
contours in the $\Delta m^2$-$\tan^2\theta$ plane. 
In fig.~[\ref{fexclusions}-(left)] we show the exclusion plots for the
solar, radiochemical + Cerenkov solar data and KamLAND with the
global signal of the SNO phase-II data, whereas the right 
panel refers to the KamLAND spectrum, radiochemical + Cerenkov solar data 
and the SNO phase-II spectrum information. Contour lines correspond to the 
the allowed areas at 90, 95, 99 and 99.7\% CL 
relative to the absolute minimum.

\begin{figure*}
\centering 
\begin{tabular}{cc} 
\includegraphics[scale = 0.6]{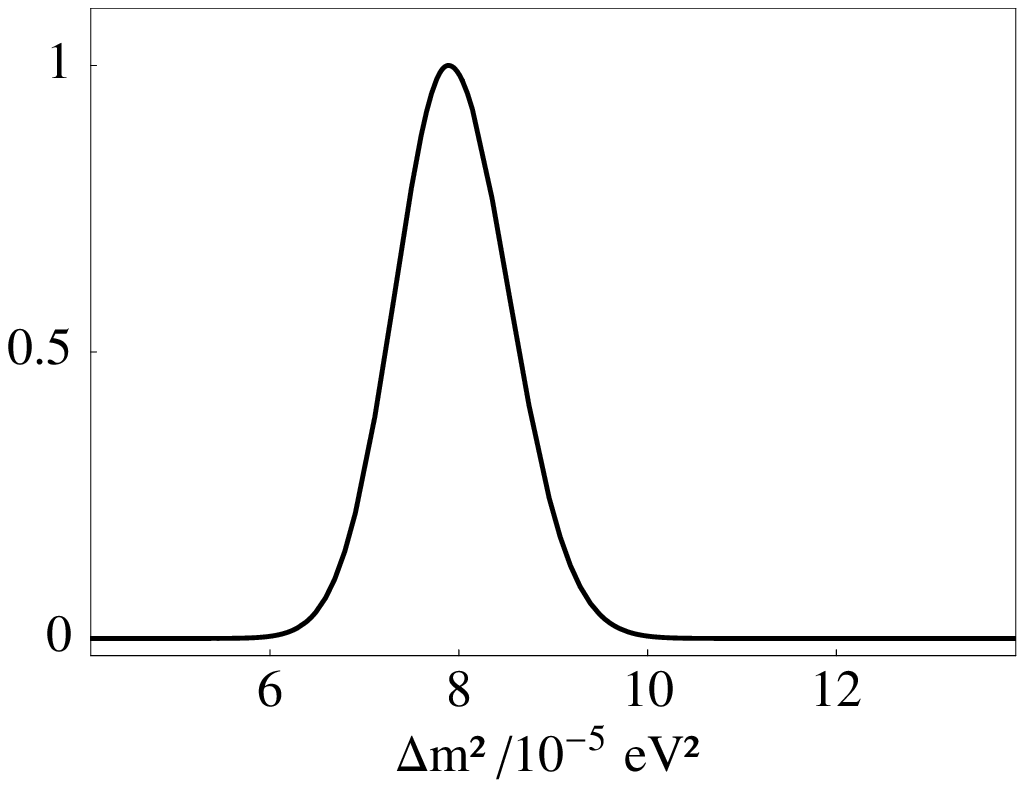}& 
\includegraphics[scale = 0.6]{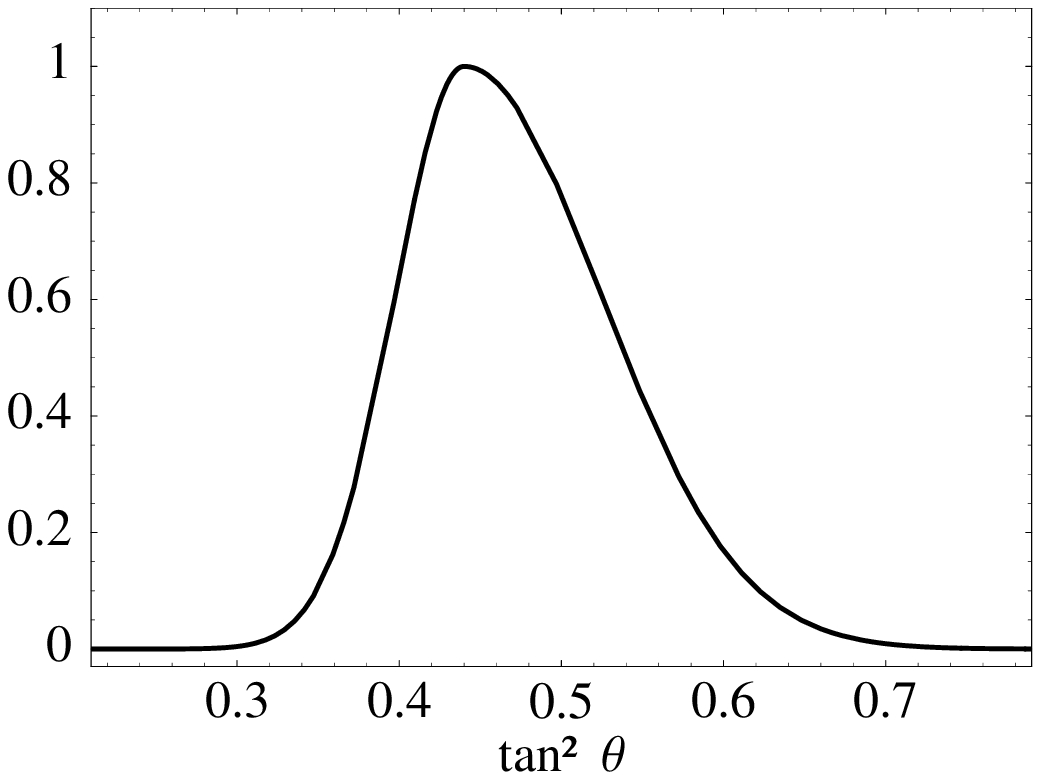} 
\end{tabular}
\caption{  
Marginalized likelihood distributions for each of the 
oscillation parameters $\Delta m^{2}_{\odot}$ (left) and  
$\tan^2\theta$ (right) corresponding to the totality 
of solar and KamLAND data. The curves are in arbitrary 
units with normalization to the maximum height. 
} 
\label{fmarginal} 
\end{figure*}

Thes normalized marginal likelihood, obtained from the integration of ${\cal L}$ for each 
of the variables, is plotted in Figs.~(\ref{fmarginal}) 
for each of the oscillation 
parameters $\Delta m^2$ and $\tan^2\theta$. 
Concrete values for the parameters are extracted by fitting  one- or two-sided 
Gaussian distributions to any of the peaks (fits not showed in the plots). In both cases, for  angle 
and the mass difference distributions the goodness of fit of the Gaussian fit to each individual peak 
is excellent (g.o.f $\sim 100\%$). 
The errors obtained from this method are assigned to the $\chi^2$ minimisation values.
The central values are fully consistent and very similar to 
the values obtained from simple $\chi^2$ minimisation. 
Systematics variability of these results can come from  the use 
of a  different prior information or mixing parameterizations, 
however this variability or systematic error due to the 
procedure is small. We will again use the technique of marginal 
distributions in the next paragraphs to obtain an estimation 
of the individual elements of the neutrino mass matrix and their 
errors.

The main difference with previous analysis is a better resolution in 
parameter space. 
The previously two well separated solutions  LMAI,LMAII have now completely 
disappeared.
In particular the secondary region at larger mass differences
(LMAII) is now completely excluded.

The introduction of the new KamLand data in general strongly 
diminishes the favored value for the 
mixing angle with respect to the KamLAND result alone \cite{Aliani:2002na}. 
The final value is  more near to those values favored by the solar data 
alone than to the KamLAND ones. 
As an important consequence, 
the combined analysis of solar and KamLAND data concludes that maximal mixing is 
not favored at $\sim 4-5\sigma$. 
This conclusion is not supported by the 
antineutrino, earth-controlled, conceptually simpler KamLAND results alone.
As we already pointed out in Ref.\cite{Aliani:2002na}, 
this effect could be simply due to the present low 
KamLAND statistics or, more worrying, to some statistical artifact derived from the complexity 
of the analysis  and of the heterogeneity of binned data involved.

\subsection{An estimation of the neutrino mass matrix}

We proceed now to an estimation of the neutrino mass matrix in
different aproximations.
Our main objective is however to  estimate how well the 
individual errors of the mass matrix can be extracted 
already at present by the existing experimental evidence.
For this purpose we have applied similar arguments as those used
before  to obtain marginal distributions and errors for individual
parameters from them. 

The square of the neutrino mass matrix can be written in the flavour basis as 
$M^2=U M_D^2 U^\dagger $
 where $M_D$ is diagonal and $U$ is an unitary (purely 
active oscillations are assumed) mixing matrix. Subtracting one of the diagonal 
entries  we have
$$ M^2=m_1^2 I+M_0^2= m_1^2 I + U M_D^{\prime 2} U^\dagger, $$
where $I$ is the identity matrix. In this way we distinguish in the mass matrix a 
part, $M_0^2$, which affects and can be determined by  oscillation experiments
and another one,  $m_1^2 I$, which does not. Evidently, 
the off-diagonal elements of the mass matrix  are fully measurable 
 by oscillation experiments.  

First, we restric ourselves for the sake of simplicity to 
two neutrino oscillations, we have in this case
{\small
\begin{eqnarray}
M^2=m_1^2 I+M_0^2&=& m_1^2 I + \Delta m^2
\pmatrix{\sin^2\theta & \sin\theta \cos\theta \cr
\sin\theta \cos\theta & \cos^2\theta }
\label{eq1000}
\end{eqnarray}
}
with $\Delta m^2=m_2^2-m_1^2$. 
The individual elements of the matrix $M_0$ can simply be
estimated from the oscillation parameters obtained before. For example for 
$\tan^2\theta\sim 0.40$, $\Delta m^2\sim 7-8\times 10^{-5}\ eV^2$ we
would obtain $(M_0^2)_{22}\sim 5-6\times 10^{-5}\ eV^2$.

Using again as likelihood  function the 
quantity \phantom{lkadlk}
${\cal L}/{\cal L}_0(\Delta m^2,\tan^2\theta) = $~$ e^{-(\chi^2-\chi_{min}^2)/2}$ 
we obtained the individual probability distributions  for any of 
the elements of the matrix $M_0$. Average values and 
$1\sigma$ errors are obtained from two-sided
Gaussian fits to these distributions.
From this procedure we obtain:  
\begin{eqnarray}
M_0^2&=& 
10^{-5}\ eV^2 
\pmatrix{ 2.05^{+0.25}_{-0.26} & 3.12^{+0.25}_{-0.26}\cr  
3.12^{+0.25}_{-0.26} & 4.50^{+0.51}_{-0.40}}.
\label{eq1001}
\end{eqnarray}

One can go further supposing a concrete value for $m_1^2$ from elsewhere. If
we take $m_1^2 >> \Delta m^2$ then we can directly write the
mass matrix
\begin{eqnarray}
M&=&m_1 I+\frac{1}{2 m_1} M_0^2.
\end{eqnarray}
Supposing for example $m_1=1\ eV$,
{\small
\begin{eqnarray*}
M&=& eV 
\pmatrix{ 1.0+1.02^{+0.12}_{-0.12}\ 10^{-5}& 1.56^{+0.12}_{-0.13}\ 
10^{-5}\cr  
1.56^{+0.12}_{-0.13}\ 10^{-5}& 1.0+2.25^{+0.25}_{-0.20}\ 10^{-5}}.
\end{eqnarray*}
}
this is  the final two neutrino mass matrix which can be 
obtained from present 
oscillation evidence coming from solar and reactor neutrinos.

We  obtain now an estimation of the three neutrino mass matrix.
For this purpose we make  the same reasoning as before and 
introduce the existing evidence of the individual values of 
the two additional angles and the square mass difference. 
Naturally knowledge of these parameters is still very poor and the 
elements of the final mass matrix will have much larger errors.
First we substract a diagonal part and write the 
square mass matrix $M^2$ as:
\begin{eqnarray*}
M^2=m_2^2 I+M_0^2&=& m_2^2 I + \Delta m^2_{12} M_{12}^2
+\Delta m^2_{32} M_{32}^2
\end{eqnarray*}
with $\Delta m^2_{ij}=m_i^2-m_j^2$. 

We write the mixing matrix as a product of three single
rotations around each of the axis:
$$U=u_{12}(\theta_{12})u_{23}(\theta_{23})u_{13}(\theta_{13}).$$
With this notation the matrices $M_{12},M_{32}$ are 
written
$$M_{12}= (u_{23} u_{13})^t M_0^2 u_{23} u_{13},$$
$$M_{32}= (u_{23} u_{13})^t M_3 u_{23} u_{13}$$
where $M_3=Diag(0,0,1)$. The matrix $M_{32}$ does not 
depend on the angle $\theta_{12}$. The dependence on 
this angle is fully contained in $M_0$ which is the 
$3\times 3$ enlarged version of the $M_0$ matrix  
appearing in Eqs.(\ref{eq1000},\ref{eq1001}).
 
We take the best values for $\Delta m_{32}$ known 
at present (see for example  Ref.\cite{neutrinos3} and references therein)
and from CHOOZ evidence \cite{chooznew,PaloVerde} 
the value of the $(13)$ angle:
$ 1.3 \,\times\, 10^{-3}\,\mbox{eV}^2\,\leq\, |\Delta m^2_{atm}|\, \leq \,3.1\, \times\,10^{-3}\,\mbox{eV}^2,$
$ 0.90 \leq \sin^22\theta_{23} \leq 1.0,$
 $\sin^2 \theta_{13} < 0.047,\ 90\%~{\rm C.L.}.$
With this values and for those 
values obtained previously for the 
$2\times 2$ $M_0^2 $ matrix (Eq.[\ref{eq1000},\ref{eq1001}])
we finally obtain an estimation for 
the three neutrino squared mass matrix ($10^{-5}\ eV^2$ units) :
{\small
\begin{eqnarray*}
M_0^2&=& 
\pmatrix{ 8.1\pm 6.5 & 8.4\pm 6.6 & -27.7\pm 28.0
\cr       8.4\pm 6.6 & 8.6\pm 6.9 & -27.5\pm 29.0
\cr       -27.7\pm 28.0 & -27.6\pm 29.0 & 202.0\pm 60.0}.
\end{eqnarray*}
}

One can go further supposing a concrete value 
for the free parameter $m_2^2$ from elsewhere. 
If we take $m_2^2 >> \Delta m^2$ then we 
can directly write the mass matrix
\begin{eqnarray}
M&=&m_2 I+\frac{1}{2 m_2} \left (M_{21}^2+M_{32}^2\right ).
\end{eqnarray}
Supposing for example $m_1=1\ eV$, we obtain ($eV$ units) 
{\small
\begin{eqnarray*}
M&=& 
\pmatrix{ 1+
(4.0\pm 3.2)\ 10^{-5}& 4.2\pm 3.2\ 10^{-5} &-13.5\pm 14.0\ 10^{-5} 
\cr  
4.2\pm 3.2\ 10^{-5}& 1+(4.3\pm 3.5)\ 10^{-5} &-13.5\pm 14.5\ 10^{-5} 
\cr 
13.5\pm 14.0\ 10^{-5}& -13.5\pm 14.5\ 10^{-5} &1+(100.0\pm 30.0)\ 10^{-5} 
}.
\end{eqnarray*}
}
this is now  the final three neutrino mass matrix which can be obtained from present 
oscillation evidence coming from solar and reactor neutrinos.

\section{Summary and  Conclusions}
\label{summary}
 
We have presented an up-to-date  analysis including the recent KamLAND
results, the SNO-phase II spectrum and all other solar neutrino data
The  active  
neutrino oscillations hypothesis has been confirmed, and the
decoupling of the atmospheric $\Delta m^2$-solar 
$\Delta m^2$ justifies a 2-flavour analysis as the one presented 
here. 
This justification is even stronger if we have into 
account the large experimental disparity among 
solar, earth reactor and atmospheric evidence and the 
very much different accuracy which can be obtained 
in each of them for the parameters of the $\mu\tau$ and 
$e\mu$ neutrino sectors.
Moreover, the consideration in the analysis
 of the atmospheric data would only slightly modify 
the best values and allowed regions 
for the parameters. These modifications would be well 
within the error bars of these paramters according to the 
present determination.

The results presented along this work show
how 
due to the increased statistics, the inclusion of the new 
KamLAND data determines with good accuracy the value 
of $\Delta m^2_{\odot}$, clearly 
selecting the LMAI solution, and brings us to a new era 
of precision measurements in the solar neutrino
parameter space \cite{kamlandTalk}.

We have introduced in this work diverse novelties in the
treatment of the SNO and Kl spectra. For the first 
one we have improved upon previous works in the 
full consideration of the sistematic correlations. 
For the KL spectrum we have studied the variability 
of the best fit results with respect the statistical 
method in use. We have shown that appreciable 
differences can be obtained.
We believe that a careful study and proper statistical 
treatment of the KL evidence is needed. Significant 
differences on the values of the oscillation parameters 
can be obtained basically due to poor statistics. 
These apparition of these differences can be easily missed 
or obscured by analysis which include large quantities 
of diverse data without the needed care of the individual 
components.

We have  obtained the allowed area in parameter space 
and individual 
values for $\Delta m^2$ and $\tan^2\theta$ with error estimation
 from the analysis of marginal likelihoods. 
We have shown that it is already possible to 
determine at present active two neutrino oscillation parameters with
relatively good accuracy.
In the framework of two active neutrino oscillations we obtain 
$$
 \Delta m^2= 8.20\pm 0.08\times 10^{-5} \eV^2,\quad \tan^2\theta=
 0.50^{+0.11}_{-0.06}.
$$ 
 We estimate the individual elements of the two neutrino mass matrix,
we show that individual elements of this matrix can be determined  
with an error $\sim 10 \% $ from present experimental evidence.

The use of the SNO phase-II spectrum in the data set has 
mainly two
effects: 1) a slight reduction in the overall area of the 
exclusion plot
and 2) a slight decrease in the best-fit $\Delta m^2_{\odot}$.

The decrease in the
best-fit mass squared difference can be understood by the fact 
that by
including the SNO spectrum, we increase the statistical 
relevance of solar
neutrino data, which {\em prefer smaller $\Delta m^2$}. 
Furthermore, the
oscillation pattern (whose information is contained in the 
spectrum) is
more sensitive to $\Delta m^2$.


It is interesting to note that the KamLAND data alone
still continue to predict, for both their analyses, a value of 
$\tan^2\theta$
 smaller than
the one obtained with the previous data, and significantly different
from 1, consequently making the aesthetically pleasing 
bi-maximal-mixing
models strongly disfavored. This result confirms 
what was already evident
in the solar neutrino data analyses. Nevertheless, 
improvement on the
determination of $\tan^2\theta$ is necessary and it is known 
that KamLAND is  
only slightly sensitive to this mixing parameter. 
The (lower) accuracy
with which we determine the solar mixing angle is evident in 
the marginalized 
likelihood plots of fig.~[\ref{fmarginal}].  
Planning of future super-beam experiments
aimed at determining the $\theta_{13}$ and eventual CP 
violating phases
relies on the most 
accurate estimation of all the mixing 
parameters~\cite{terranova}. 
It is expected that future solar neutrino experiments, 
notably phase-II SNO
(higher statistics, due to be made public soon)
and eventually future low energy experiments, and phase-III
SNO (with helium) will further restrict the allowed 
range of parameters.

\vspace{0.3cm}
\subsection*{Acknowledgments}

We would like to thank F. Terranova 
and M. Smy for usefull discussions. 
We  acknowledge the  financial  support of 
 the Italian MIUR, the  Spanish CYCIT  funding agencies 
 and the CERN Theoretical Division.
P.A. acknowledges funding from the Inter-University Attraction Pole (IUAP) 
"fundamental interactions".
The core of the numerical computations were done at the  
computer farm of the Universit\`a degli Studi di Milano, Italy.

\newpage

\begin{table*}
\parbox{\linewidth}{
\centering
\begin{tabular}{lcl} 
Bin (MeV) & $S_{exp}$/MC & $\sigma_{stat.}$ \\ \hline 
2.600 - 3.025 & 0.45 & 1.4\\ 
3.025 - 3.450 & 0.56 & 1.5\\ 
3.450 - 3.875 & 0.67 & 1.7\\ 
3.875 - 4.300 & 0.62 & 2.1\\ 
4.300 - 4.725 & 0.99 & 2.6\\ 
4.725 - 5.150 & 1.20 & 3.3\\ 
5.150 - 5.575 & 0.80 & 4.5\\ 
5.575 - 6.000 & 1.00 & 6.7\\ 
6.000 - 6.425 & 1.20 & 10.0\\ 
6.425 - 6.850 & 0.33 & 17.0\\ 
6.850 - 7.275 & 0.67 & 33.0\\ 
7.275 - 7.700 & 0.00 & - \\ 
7.700 - 8.125 &   -  & - \\
\hline 
\end{tabular} 
}
\caption{Summary of Kamland spectrum 
information extracted  from \cite{klJun}. 
Relative statystical errors only 
are reported.} 
\end{table*}

\begin{table*}
\centering
\begin{tabular}{cc|cc} 
\hline 
$T_{eff}$ (MeV)& Evnts./500 keV &$T_{eff}$ (MeV)& Evnts./500 keV\\ 
\hline 
 5.5- 6.0    & 225 & 9.5-10.0    & 155\\ 
 6.0- 6.5    & 225 &10.0-10.5    &  80\\ 
 6.5- 7.0    & 220 &10.5-11.0    &  95\\ 
 7.0- 7.5    & 255 &11.0-11.5    &  55\\ 
 7.5- 8.0    & 235 &11.5-12.0    &  40\\ 
 8.0- 8.5    & 225 &12.5-13.0    &  15\\   
 8.5- 9.0    & 155 &13.5-14.0    &  15\\   
 9.0- 9.5    & 145 &14.5-15.0    &  5\\   
\\ \hline 
\end{tabular}

\begin{tabular}{cc|cc} 
\hline 
$T_{eff}$ (MeV)& Evnts./500 keV &$T_{eff}$ (MeV)& Evnts./500 keV\\ 
\hline 
 5.5- 6.0    & $840\pm 20$  & 9.5-10.0  & $180    \pm 10   $ \\ 
 6.0- 6.5    & $785\pm 20$  &10.0-10.5  & $110    \pm 10   $ \\ 
 6.5- 7.0    & $705\pm 20$  &10.5-11.0  & $110    \pm 10   $ \\ 
 7.0- 7.5    & $680\pm 20$  &11.0-11.5  & $50    \pm  10  $ \\ 
 7.5- 8.0    & $560\pm 15$  &11.5-12.0  & $40    \pm  10  $ \\ 
 8.0- 8.5    & $470\pm 15$  &12.5-13.0  & $10    \pm  10  $  \\   
 8.5- 9.0    & $245\pm 10$  &13.5-14.0  & $10    \pm  5  $ \\   
 9.0- 9.5    & $205\pm 10$  &14.5-15.0  & $0    \pm  5  $ \\   
\\ \hline 
\end{tabular} 
\caption{ (top) CC Energy spectrum observed at SNO 
(taken from fig.~36 of ref~\cite{newSNO}).
(bottom) SNO neutrino spectrum (
fig.24 from Ref.\cite{newSNO}.
}
\label{sno_spectrum} 
\end{table*}

\begin{table*}
\centering
\parbox{\linewidth}{
\centering
\begin{tabular}{lcr} 
  Experiment &&References\\ \hline 
  Homestake& &\protect\cite{homestake}\\ 
  SAGE & &\protect\cite{sage1999,sage}\\ 
  GallEx &&\protect\cite{gallex}\\ 
  GNO &&\protect\cite{gno2000}\\ 
SuperKamiokande & &\protect\cite{Smy:2002fs}\\ 
SNO & & \protect\cite{newSNO,Poon:2001ee,Ahmad:2002jz}\\ 
CHOOZ && \protect\cite{chooznew} \\ 
Palo Verde & & \protect\cite{PaloVerde}\\ 
KamLAND & &\protect\cite{klJun,Eguchi:2002dm}\\  
\hline 
\end{tabular} 
}
\caption{References from where we draw the data used in our analysis.} 
\end{table*}

\begin{table*}
\centering
\begin{tabular}{lll} 
& $\Delta m^2 (\eV^2) $& $\tan^2\theta$\\ \hline
{from $\chi^2$ minimization} & & \\
\hspace{1cm} 
KL (Sp+Gl)+Solar + SNO (Sp)&  $7.89\times 10^{-5}$ & $0.40$ \\
\hspace{1cm}  
KL (Sp+Gl)+Solar + SNO (Gl)&  $8.17\times 10^{-5}$ & $0.40$ \\ 
{from marginalization} & & \\
\hspace{1cm}  
KL (Sp+Gl)+Solar + SNO (Sp)&  $8.2^{+0.9}_{-0.8}\times 10^{-5}$ & $0.50^{+0.11}_{-0.06}$ \\ 
\hline 
\end{tabular} 
\caption{ Mixing parameters  from $\chi^2$ minimization 
and likelihood marginalization. }
\label{bestfitpoints} 
\end{table*}

\end{document}